\documentclass[aps,prb,twocolumn,superscriptaddress,floatfix,10pt]{revtex4-1}

\usepackage{graphicx}
\usepackage{epstopdf}
\usepackage{color}
\usepackage{ifsym}
\usepackage{amssymb}
\usepackage{bm}
\usepackage{upgreek}

\usepackage{lettrine}
\usepackage{type1cm}

\usepackage[bf, sf, justification=raggedright]{caption}
\DeclareCaptionLabelSeparator{line}{ $|$ }
\usepackage{xr}
\newcommand{\degrees}{$^{\circ}$}

\begin{document}

\title{\textsf{Synthetic ferrimagnet nanowires with very low critical current density\\ for coupled domain wall motion}}

\author{Serban~Lepadatu}
\affiliation{School of Physics \&\ Astronomy, University of Leeds, Leeds LS2 9JT, United Kingdom}
\affiliation{Jeremiah Horrocks Institute for Mathematics, Physics \&\ Astronomy, University of Central Lancashire, Preston, Lancashire PR1 2HE, United Kingdom }

\author{Henri~Saarikoski}
\affiliation{RIKEN Center for Emergent Matter Science (CEMS), 2-1 Hirosawa, Wako, Saitama, 351-0198 Japan}

\author{Robert~Beacham}
\affiliation{Scottish Universities Physics Alliance, School of Physics \&\ Astronomy, University of Glasgow, Glasgow G12 8QQ, United Kingdom}

\author{Maria~Jose~Benitez}
\affiliation{Scottish Universities Physics Alliance, School of Physics \&\ Astronomy, University of Glasgow, Glasgow G12 8QQ, United Kingdom}
\affiliation{Departamento de F\'{\i}sica, Escuela Polit\'{e}cnica Nacional, Quito, Ecuador}

\author{Thomas~A.~Moore}
\affiliation{School of Physics \&\ Astronomy, University of Leeds, Leeds LS2 9JT, United Kingdom}

\author{Gavin~Burnell}
\affiliation{School of Physics \&\ Astronomy, University of Leeds, Leeds LS2 9JT, United Kingdom}

\author{Satoshi~Sugimoto}
\affiliation{School of Physics \&\ Astronomy, University of Leeds, Leeds LS2 9JT, United Kingdom}

\author{Daniel~Yesudas}
\affiliation{School of Physics \&\ Astronomy, University of Leeds, Leeds LS2 9JT, United Kingdom}

\author{May~C.~Wheeler}
\affiliation{School of Physics \&\ Astronomy, University of Leeds, Leeds LS2 9JT, United Kingdom}

\author{Jorge~Miguel}
\affiliation{Diamond Light Source, Chilton, Didcot OX11 0DE, United Kingdom}

\author{Sarnjeet~S.~Dhesi}
\affiliation{Diamond Light Source, Chilton, Didcot OX11 0DE, United Kingdom}

\author{Damien~McGrouther}
\affiliation{Scottish Universities Physics Alliance, School of Physics \&\ Astronomy, University of Glasgow, Glasgow G12 8QQ, United Kingdom}

\author{Stephen~McVitie}
\affiliation{Scottish Universities Physics Alliance, School of Physics \&\ Astronomy, University of Glasgow, Glasgow G12 8QQ, United Kingdom}

\author{Gen~Tatara}
\affiliation{RIKEN Center for Emergent Matter Science (CEMS), 2-1 Hirosawa, Wako, Saitama, 351-0198 Japan}

\author{Christopher~H.~Marrows}\email[email:~]{c.h.marrows@leeds.ac.uk}
\affiliation{School of Physics \&\ Astronomy, University of Leeds, Leeds LS2 9JT, United Kingdom}

\begin{abstract}
Domain walls in ferromagnetic nanowires are potential building-blocks of future technologies such as racetrack memories, in which data encoded in the domain walls are transported using spin-polarised currents. However, the development of energy-efficient devices has been hampered by the high current densities needed to initiate domain wall motion. We show here that a remarkable reduction in the critical current density can be achieved for in-plane magnetised coupled domain walls in CoFe/Ru/CoFe synthetic ferrimagnet tracks. The antiferromagnetic exchange coupling between the layers leads to simple N\'{e}el wall structures, imaged using photoemission electron and Lorentz transmission electron microscopy, with a width of only $\sim 100$~nm. The measured critical current density to set these walls in motion, detected using magnetotransport measurements, is $1.0 \times 10^{11}$~Am$^{-2}$, almost an order of magnitude lower than in a ferromagnetically coupled control sample. Theoretical modelling indicates that this is due to nonadiabatic driving of anisotropically coupled walls, a mechanism that can be used to design efficient domain-wall devices.
\end{abstract}


\date{\today}
\maketitle

\lettrine[lines=3]{\textcolor[gray]{0.5}T}{}he presence of antiferromagnetic\cite{Gruenberg1986} and oscillatory\cite{Parkin1990} indirect exchange coupling across a metal spacer between ferromagnetic layers was one of the earliest discoveries in the field of ultrathin film magnetism. The effect can be used to construct synthetic antiferromagnets (SAFs), where the alternating atomic spins are replaced by the alternating moments of magnetic layers\cite{Steadman2002}. These structures have found applications as the reference layers in magnetoresistive recording heads\cite{Pinarbasi1999}, and in the operation of toggle-mode magnetic random access memory cells\cite{Slaughter2005}.

Domain walls (DWs) in magnetic nanowires\cite{Boulle2011} have been proposed as the basis of memory\cite{ParkinRacetrack2015}, logic\cite{Allwood2005,Xu2008}, and sensor\cite{Diegel2009} technologies. By far the most common choice of material in the past has been Permalloy (Ni$_{80}$Fe$_{20}$)\cite{yamaguchiprl2004,Beach2006,Hayashi2007b,Meier2007,Lepadatu2009a}. Shape anisotropy means that the domains in such a soft magnetic nanowire have magnetisation that points along the wire length, leading to head-to-head or tail-to-tail DWs. The micromagnetics of these ribbon-shaped wires is such that the walls take non-trivial vortex or transverse forms\cite{McMichael1997}. The width of the DWs is tied to the wire geometry\cite{nakatanijmmm2005}, with very high current densities required to set them in motion using spin-transfer torques\cite{Locatelli2014}, typically a few $10^{12}$~Am$^{-2}~$ (Refs~ \cite{yamaguchiprl2004,Meier2007,Lepadatu2009b,Heyne2009}). Whilst moving, neither type of wall is a rigid object, since they possess internal degrees of freedom that can dissipate energy\cite{Klaui2005}. The stray field from the DWs means that they will also couple if brought too close together\cite{Parkin2008}. These drawbacks are all major obstacles to the use of DWs in such nanowires in applications such as racetrack memories\cite{ParkinRacetrack2015}.

One approach to move towards more useful devices is to employ chiral walls\cite{Thiaville_DMI} in ultrathin perpendicularly magnetised layers that possess a Dzyaloshinskii-Moriya interaction. Such walls are narrow, with a simple N\'{e}el structure, and couple well to spin-orbit torques\cite{emori2013current,ryu2013chiral}, but still
interact with each other by their stray fields. Perpendicularly magnetised SAF wires do not have this drawback, and show very high wall velocities when driven with high current densities but still require at least $\sim 10^{12}$~Am$^{-2}$ current density for the onset of wall motion\cite{Yang2015}.

A structure with unbalanced antiparallel moments is a synthetic ferrimagnet (SyF). Here we show that returning to the use of in-plane magnetised nanowires, formed into SyFs, can mitigate against all of these obstacles. In particular, the threshold current density can be significantly reduced when the combination of isotropic indirect exchange across the Ru spacer and anisotropic magnetostatic coupling between the walls allows the engineering of a situation where the walls are coupled antiferromagnetically in-plane, but ferromagnetically out-of-plane. Within a one-dimensional (1-D) model\cite{Saarikoski2014}, we show that if this out-of-plane ferromagnetic coupling is sufficiently strong, the new internal degrees of freedom in the coupled pair of walls can be exploited to reduce the threshold current density to zero, even in the presence of finite extrinsic pinning. This is due to the coupled wall pair possessing internal degrees of freedom that a wall in single wire does not, which can be exploited to escape the pinning potential. We have tested this experimentally in a CoFe/Ru/CoFe SyF nanowire: measurements of the current-driven domain wall mobility curve in zero magnetic field have shown critical current densities as low as $J_\mathrm{crit} \approx 1.0 \times 10^{11}$~Am$^{-2}$. This is less than half the lowest reported critical current density to date, $\sim 2.5 \times 10^{11}$~A/m$^2$, measured in a Co/Ni multilayer wire in the pure intrinsic pinning regime \cite{Koyama2011}. Our work expands the range of possible designs available to engineers when trading off parameters such as speed, density, and power consumption when designing domain wall devices. Such structures are thus good candidates for racetrack memories where power consumption is the most important figure-of-merit.

\section*{Results}

\begin{figure*}[t]
  \begin{center}
    \includegraphics[width=14cm]{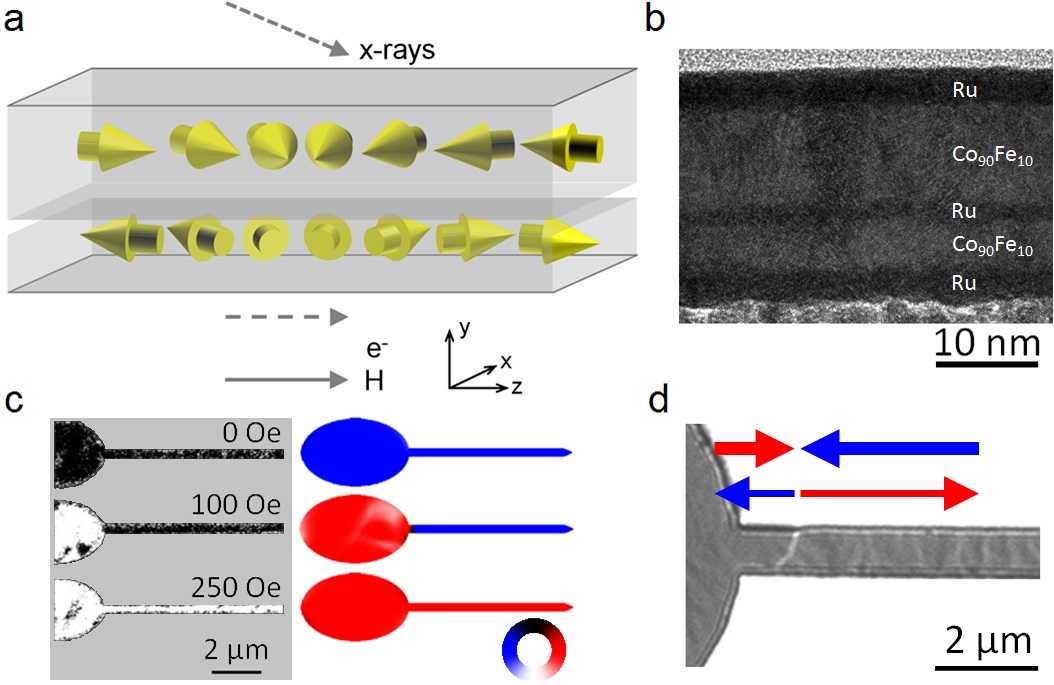}
  \end{center}
  \caption{\textsf{\textbf{Domain walls in synthetic ferrimagnet nanostructures.} \textbf{a,} Diagram of SyF track indicating the directions of magnetisation, x-ray beam in the PEEM, electron flow (e$^-$) during magnetotransport, applied field ($H$) and defined coordinate system. \textbf{b,} High-resolution TEM cross-section of SyF thin film. The layers are $t_1 = 13.3$~nm and $t_2 = 6.6$~nm thick separated by a $0.7$~nm thick Ru spacer layer. \textbf{c,} XMCD-PEEM images and micromagnetic simulations of SyF structure with 400~nm wide track, showing reversal of ellipse and track in the thicker Co$_{90}$Fe$_{10}$ layer. XMCD-PEEM images are taken at remanence after application of a field pulse, with contrast sraing only from the upper, thicker layer. The color wheel indicates the magnetisation directions in the same layer in the micromagnetic simulations. \textbf{d,} LTEM image of injected DW in SyF track with 800~nm width. The measured DW width is $\sim 100$~nm. The red and blue arrows indicate the inferred direction of magnetisation in the two layers.} \label{fig:SAFstructure}}
\end{figure*}

\subsection*{Structure of coupled domain wall pairs in SyFs}

In order to study the DW structures in these trilayers, SyF nanowire devices were fabricated as described in the Methods section with the geometry shown in Fig.~\ref{fig:SAFstructure}c: a 16~$\upmu$m long, 400~nm wide nanowire with one pointed end and the other attached to a large elliptical injection pad \cite{Lepadatu2009a}. The SyF comprised a Co$_{90}$Fe$_{10}$~($t_2$)/Ru/Co$_{90}$Fe$_{10}$~($t_1$) trilayer. The high-resolution TEM cross-section in Fig.~\ref{fig:SAFstructure}b clearly resolves the Ru spacer layer, the 0.7~nm thickness of which is chosen to give strong antiferromagnetic coupling of the two Co$_{90}$Fe$_{10}$ layers\cite{Parkin1990}, which have thicknesses $t_1 = 13.3$ and $t_2 = 6.6$~nm.  We chose to study the unbalanced SyF configuration since magnetic fields may be used to initiate the devices in a controlled magnetisation configuration, due to the fact that the net magnetic moment is not fully compensated.

By analysing the magnetostatics in SyFs it may be shown that elliptical elements have a significantly lower coercivity than rectangular segments\cite{Fruchart2012}, just as for single ferromagnetic layers\cite{Schrefl1997}. Coupled pairs of DWs--shown schematically in Fig.~\ref{fig:SAFstructure}a--were prepared in the SyF nanostructures by applying a controlled reverse magnetic field to switch the net magnetisation in the elliptical element. This process is shown in Fig.~\ref{fig:SAFstructure}c using X-ray magnetic circular dichroism photo-emission electron microscopy (XMCD-PEEM) imaging, accompanied by micromagnetic simulations that reproduce the injection process. The XMCD-PEEM technique is a surface sensitive method, thus the images in Fig.~\ref{fig:SAFstructure}c represent the magnetisation direction only in the upper, thicker, Co$_{90}$Fe$_{10}$ layer. Initially the devices are saturated with net magnetisation pointing to the left, driven by the magnetisation in the thicker layer responding to the applied field. Longitudinal field pulses up to 100~Oe reverse the net magnetisation in the elliptical element whilst leaving the magnetisation in the nanowire unchanged, generating a coupled pair of DWs at the neck of the elliptical pad. Stronger field pulses propagate the DW along the track, eventually reversing its net magnetisation direction.

\begin{figure*}
  \begin{center}
    \includegraphics[width=12cm]{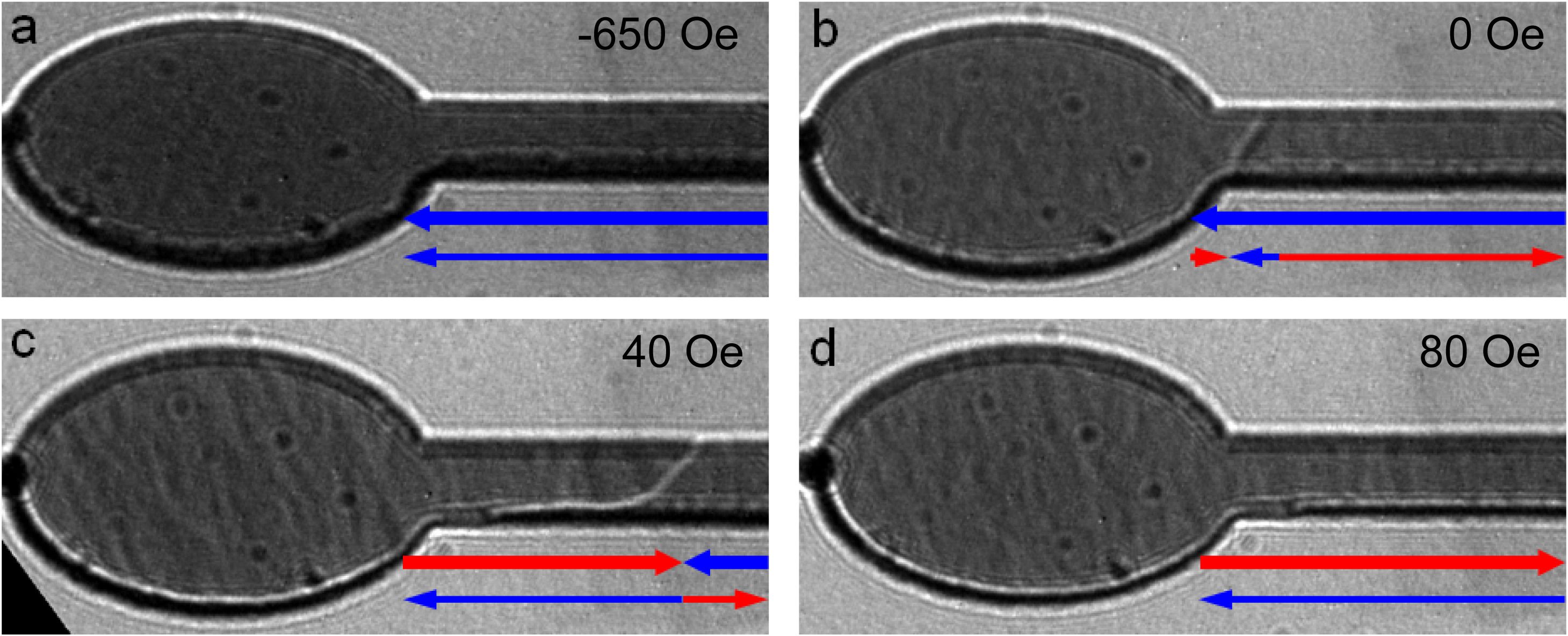}
  \end{center}
  \caption{\textsf{\textbf{LTEM images of SyF nanowire with 800~nm wide track.} The sequence of images shows the reversal of the ellipse and nucleation of a 180\degrees~AFM DW. The layers are $t_1 = 13.3$~nm and $t_2 = 6.6$~nm thick separated by a $0.7$~nm thick Ru spacer layer. The transmission images contain contrast that arises as a projection of both layers. The images are taken for applied in-plane horizontal fields of \textbf{a,} -650~Oe, saturation state, \textbf{b,} 0~Oe, ellipse reversed and 360\degrees~wall formed in the thinner Co$_{90}$Fe$_{10}$ layer, \textbf{c,} 40~Oe, 180\degrees~AFM DW formed, and \textbf{d,} 80~Oe, AFM DW removed. The red and blue arrows indicate the inferred direction of the magnetisation in each layer, with the thick/thin arrows representing the thicker (top) and thinner (bottom) layer respectively.} \label{fig:SAF_LTEM}}
\end{figure*}

A more detailed view of an example of the novel type of DW observed in these structures is shown in the Lorentz transmission electron microscopy (LTEM) image of an 800~nm wide wire and pad shown in Fig.~\ref{fig:SAFstructure}d. This image was taken in the defocussed Fresnel mode which is sensitive to gradients in magnetic induction, such as DWs\cite{benitez2015}. We note that in this case the contrast arises from the projected sum of in-plane magnetic induction contributions from both layers. Note that patterned films where the magnetisation is parallel to the edge of the structure also show black/white contrast due to the diverging/converging effect of the electron beam either side of the edge in Fresnel mode. Thus we can observe a white DW running across the wire in Fig.~\ref{fig:SAFstructure}d, but in addition we also see a significant change in contrast on the upper edge of the wire either side of the DW. The latter signifies a change of the net magnetic induction either side of this DW.

In order to verify the antiparallel alignment of the Co$_{90}$Fe$_{10}$ layers we have investigated the reversal and DW injection mechanism using LTEM in this 800~nm wide wire in greater detail. In the magnetically saturated state shown in Fig.~\ref{fig:SAF_LTEM}a, no DW contrast is visible in the wire itself and here the dark and bright contrast from the opposite edges shows that a net magnetisation exists in the wire which corresponds to both layers having parallel magnetisation. (This is in addition to the normal Fresnel contrast associated with the phase change at an edge which is also visible as a black/white contribution on each edge.)

On reduction of the field a 360\degrees~DW appears close to the neck of the wire and ellipse region, shown in Fig.~\ref{fig:SAF_LTEM}b. Considering the reversal mode of the SyF structure, where switching into a largely antiparallel state is expected, and comparison with the micromagnetic simulations in the Supplementary information indicates that the 360\degrees~DW exists only in the thinner layer, whilst the thicker layer remains uniformly magnetised. Compared to the previous image the magnetic contribution to the edge contrast has been reduced significantly, most notably with the dark contrast visible at the lower edge. This is a consequence of the fact that whilst the majority of the magnetisation remains along the wire axis, the direction in each layer is antiparallel in the domains to either side of the 360\degrees~DW, so that the net magnetisation that causes the contrast is smaller than Fig.~\ref{fig:SAF_LTEM}a. The 360\degrees~DW traverses the width of the wire, showing a distinct black/white character indicative of strong transverse magnetisation at the centre of the wall. Furthermore the magnetic contrast associated with the nanowire edges does not change either side of the wall, which is consistent with the wall being only in one layer therefore the net magnetisation either side of the wall is the same.

Further variation of the field sees the injection of a coupled pair of 180\degrees~DWs, one in each layer lying on top of each other, as shown in Fig.~\ref{fig:SAF_LTEM}c, which come to rest a short distance from the neck. Again, the coupled walls traverse the width of the wire. The edge contrast can again be seen to be much weaker than in Fig.~\ref{fig:SAF_LTEM}a, the parallel arrangement, and in fact is similar to the antiparallel arrangement in Fig.~\ref{fig:SAF_LTEM}b. However the edge contrast can clearly be seen to change either side of the wall. This is most apparent in the upper edge, which is darker on the left hand side of the wall compared to right hand side. This is consistent with an antiparallel arrangement on each side of the coupled DW pair, but with a net magnetisation in the opposite direction for each side, reflecting the imbalance of the layer thickness. This analysis also holds for the image presented in Fig.~\ref{fig:SAFstructure}d, and we can see that there is no overlapping parallel region in either case\cite{Hellwig}. This confirms that there truly is a pair of AFM-coupled 180\degrees~DWs in both cases. The coupled pair of DWs is eliminated by a further increase in the field as shown in Fig.~\ref{fig:SAF_LTEM}d. Here the weak magnetic edge contrast is consistent with a fully antiparallel alignment of the layers. In fact the antiparallel configurations in Figs.~\ref{fig:SAF_LTEM}b and~\ref{fig:SAF_LTEM}d are seen to have a net alignment in different directions from the change in upper edge contrast, this can be seen from the schematic arrows indicated for each layer underneath the figures. Note that the fact that the edge contrast between these two figures is not exactly equal and opposite reflects a degree of strucutural edge asymmetry due to lithographic flagging that is not exactly the same on both wire edges.

The high degree of flux closure in the SyF trilayer entirely changes the micromagnetics of the DWs meaning that the usual vortex/transverse wall picture\cite{McMichael1997} for single nanowires is substantially changed. Instead, these walls have a simple N\'{e}el form, with the wall widths measured from the Fresnel images in Figs~\ref{fig:SAF_LTEM}b and \ref{fig:SAF_LTEM}c being $110 \pm 15$~ nm: the width is also found to be independent of the wire geometry due to the changed magnetostatics. It should be noted that the widths measured here by the defocussed Fresnel method correspond to an upper limit for the widths of the domain walls. Micromagnetic simulations of DW structures in SAF and SyF Co$_{90}$Fe$_{10}$ tracks, with track widths from as small as 40~nm up to 1.2~$\upmu$m, have shown that narrow symmetric transversely magnetised N\'{e}el walls, with width $\sim 100$~nm, persist, in marked contrast to single layer ferromagnetic tracks\cite{benitez2015a}. This is due to the fact that the transverse demagnetising field, normally responsible for deforming symmetric transverse walls into asymmetric and finally into vortex walls in simple ferromagnetic tracks as the width is increased, can instead close a significant portion of its flux with the oppositely directed demagnetising field in the other layer of the SyF (see Supplementary Information).

\subsection*{Current-driven dynamics of coupled walls in SyFs}

\begin{figure*}[t]
  \begin{center}
    \includegraphics[width=10cm]{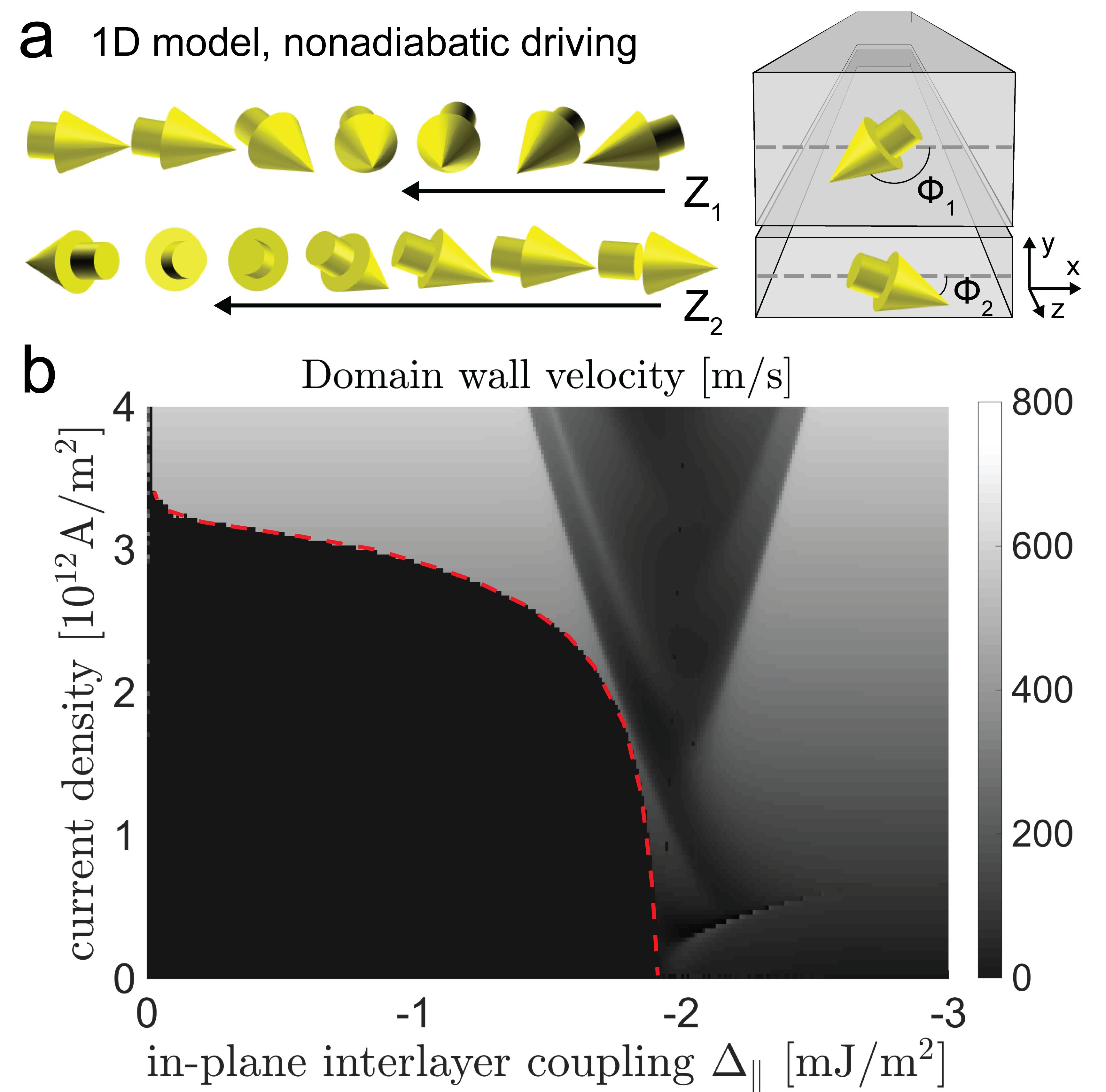}
  \end{center}
  \caption{\textsf{\textbf{Nonadiabatic driving of anisotropically coupled DWs in a SyF nanowire.} \textbf{a,} Nonadiabatic driving pushes spins out of the wire plane in the direction where the perpendicular interlayer coupling is ferromagnetic. The angles of the spins with respect to the wire plane are denoted $\phi_i$, $i=1,2$ and the positions of the walls in the 1-D model are $Z_i$, $i=1,2$. \textbf{b,} DW velocity in the 1-D model as a function of in-plane interlayer coupling $\Delta_\parallel=-(1/4)\Delta_\perp$ and driving current density, $J$, in the wire. The red dashed line indicates the threshold current for the DW motion. Walker breakdown occurs in the triangular region in the center.} \label{fig:NONADIABATIC_DRIVING}}
\end{figure*}

The current-driven dynamics of a DW may be described using a simple 1-D model\cite{tataraprl2004} that describes the DW in terms of a position co-ordinate $Z$ and a canting angle $\phi$ that acts as a conjugate momentum to $Z$. This may be readily extended to a pair of such walls (with co-ordinates $\{Z_1,\phi_1\}$ and $\{Z_2,\phi_2\}$ [see  Fig.~\ref{fig:NONADIABATIC_DRIVING}a]), that are coupled together by isotropic antiferromagnetic exchange and anisotropic magnetostatic interactions\cite{Saarikoski2014}. Whilst such models have been used in the past to describe vortex and transverse walls in single wires, a 1-D model of a 2-D wall neglects important degrees of freedom\cite{Beach2008,Yoshimura2015}. The simple, rigid walls in our SyFs means that the 1-D model is expected to give more accurate insights into the DW dynamics, although the purpose of using this analytical model is to achieve insight into the depinning mechanism, rather than to provide an exact quantitative description of the system for which numerical micromagnetic simulations incorporating full knowledge of the extrinsic defects in the sample are needed.

We employed 1-D simulations of DW dynamics to gain these insights into the threshold current density in our in-plane magnetised SyF device (the equations of motion of this system are described in the Supplementary Information). It is typical of common ferromagnetic metals that $\beta/\alpha > 1$ (Refs \onlinecite{Lepadatu2010,Sekiguchi2012}), where $\beta$ is the nonadiabatic spin torque coefficient \cite{thiavilleepl2005} and $\alpha$ is the Gilbert damping constant, and so we assume that the dynamics is in the nonadiabatic regime (see Fig.~\ref{fig:NONADIABATIC_DRIVING}a). The estimated exchange coupling and the magnetostatic couplings (see Methods section) indicate also a highly anisotropic total interlayer coupling which is ferromagnetic in the out-of-plane direction and much stronger than the antiferromagnetic in-plane coupling.

Fig.~\ref{fig:NONADIABATIC_DRIVING}b shows DW velocity as a function of interlayer coupling and driving current calculated using the 1-D model in the nonadiabatic regime (see Methods section). The anisotropic coupling between the walls, arising from a combination of interlayer exchange and magnetostatics, is parameterised by in-plane $\Delta_\parallel$ and out-of-plane $\Delta_\perp$ coupling constants.  The threshold current, the minimum current needed to depin the walls from the pinning potential and set them moving, is indicated with the red dashed line. The threshold current decreases with increasing interlayer coupling and eventually disappears completely at a strong coupling of $\Delta_\parallel \approx -1.9$~mJ/m$^2$. This can be qualitatively understood from the presence of a nonadiabatic torque which forces the spins out-of-plane\cite{tataraprl2004,tatarareview}, a phenomenon caused by spin relaxation\cite{thiavilleepl2005,zhangprl2004} (see Fig.~\ref{fig:NONADIABATIC_DRIVING}a and the Supplementary Information). This leads to a decrease in interlayer coupling energy at the center of the wall due to strong ferromagnetic out-of-plane coupling and helps free the walls from the pinning potential.

Alternatively, the reduction of threshold current can be explained using wall momentum. For a strong ferromagnetic out-of-plane coupling, the spins at the center of the walls are close to parallel (see Fig.~S3 in the Supplementary Information) and therefore  $\phi_- \equiv {1 \over 2}(\phi_1-\phi_2)$ is small. Since $\phi_-$ is the canonical momentum of the average wall position $Z_+ \equiv \frac{1}{2}(Z_1+Z_2)$, the torque and force induced by the applied current are distributed to the dynamics of $Z_+$ and $\phi_-$. If $\phi_-$ is kept small due to the nonadiabatic torque and ferromagnetic interlayer coupling, the torque and force are efficiently transferred to the $Z_+$ mode, \textit{i.e.}, to the displacement of the wall, resulting in a reduction of the threshold current. This reduction mechanism is consistent also with the fact that no significant reduction arises when the out-of-plane coupling is antiferromagnetic\cite{Saarikoski2014}. Our 1-D simulations indicate that the threshold current vanishes when the out-of-plane ferromagnetic interlayer coupling is of the same order as the out-of-plane shape anisotropy, as can be seen from the calculations where we adjusted the ratio of $\Delta_{\perp}/\Delta_{||}$ that are presented in the Supplementary Information in Fig.~S5. We conclude that careful optimisation of shape and interlayer coupling anisotropies, by taking into account the finite wire width and other 2-D effects, can be used to design highly energy-efficient DW devices. Careful choice of the layer thicknesses, explored using 1-D model simulations in the Supplementary Information, can also be used to engineer desirable interlayer coupling properties.

The interlayer coupling also affects the Walker breakdown mechanism. The regime after the Walker breakdown is visible in the central part of Fig.~\ref{fig:NONADIABATIC_DRIVING}b as a triangular region, where the dependence on the driving current is nonlinear.

\begin{figure*}
  \begin{center}
    \includegraphics[width=14cm]{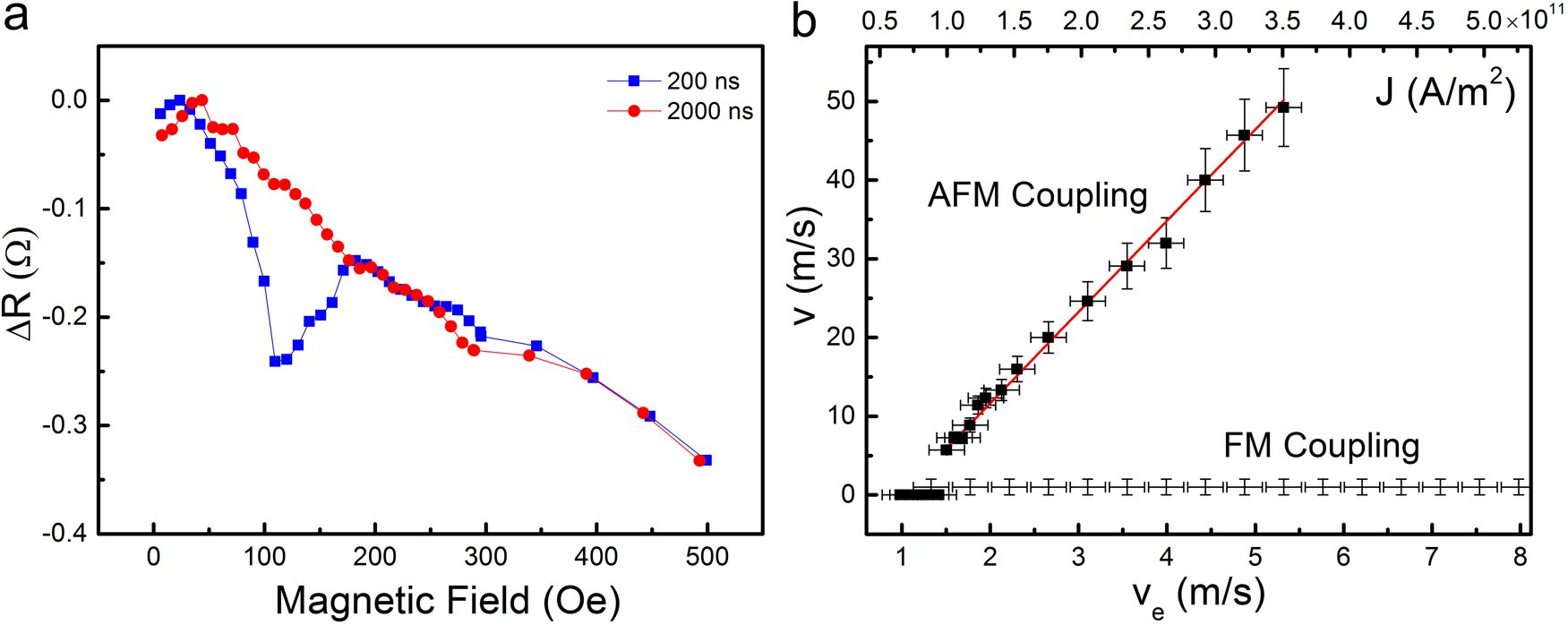}
  \end{center}
  \caption{\textsf{\textbf{Effects of pulsed currents.} \textbf{a,} Pulsed current measurements for fixed current density  and two different pulse durations showing current-driven DW speed measurement method. Short pulse duration, 200~ns, wire remains in the same magnetisation configuration. Long pulse duration, 2000~ns, wire magnetisation is fully switched and 180\degrees~DW removed. The pulse current density is $1.2 \times 10^{11}$~Am$^{-2}$. \textbf{b,} Current-driven DW mobility curve for AFM coupling obtained by determining the minimum average pulse duration required to fully remove the 180\degrees~DW between the measurement contacts for each current density value. The continuous line is a linear regression fit to the data points for values of $v_\mathrm{e}$ above the threshold where the DWs are set in motion. Results obtained from FM coupled control samples are shown for comparison.} \label{fig:magnetotransport}}
\end{figure*}

To experimentally investigate this predicted reduction in depinning current density, electrical contacts were patterned on top of the nanowire and elliptical pad of a SyF device with a 400~nm wide track and Co$_{90}$Fe$_{10}$ layers that have thicknesses $t_1 = 13.3$~nm and $t_2 = 6.6$~nm. The procedure used for measuring the current-driven mobility curve $v(v_\mathrm{e})$ of  the 180\degrees~coupled DW pair is outlined in Fig.~\ref{fig:magnetotransport}a. Here, $v$ is the DW velocity and $v_\mathrm{e}$ is the electron spin drift velocity, which is proportional to the current density (see Methods). First the coupled 180\degrees~DW pair is injected into the neck of the wire by applying a series of fields and current pulses as outlined in the Supplementary Information. This prepares a state similar to that shown in Fig.~\ref{fig:SAFstructure}d that is stable in zero field. A single current pulse of selected amplitude and duration is then applied, with electron flow away from the elliptical pad. The field, applied along the wire axis, is then swept up to saturation and the magnetoresistance curve measured. For a given current density, if the pulse duration is too short to fully remove the 180\degrees~DW from the track, the magnetoresistance shows a trough in the measured curve, a signature that the wall is still present (see Supplementary Information). An example of this for a 200~ns long pulse of $–1.2 \times 10^{11}$~Am$^{-2}$ current density is shown in Fig.~\ref{fig:magnetotransport}a. On the other hand, if the current pulse is long enough to fully drive the DW out of the track, the magnetoresistance signature does not show the characteristic trough, but decreases monotonically as the thinner Co$_{90}$Fe$_{10}$ layer is gradually rotated by the increasing magnetic field. An example of this behaviour, shown in Fig.~\ref{fig:magnetotransport}a is for a 2000~ns long pulse of $–1.2 \times 10^{11}$~Am$^{-2}$ current density.

For each current density value and pulse duration $t$, these measurements were repeated five times and the average minimum pulse length required to fully switch the magnetisation in the track was used to obtain the average current-driven DW speed $v=l/t$ using the known track length of $l = 16$~$\upmu$m. The measured mobility curve $v(v_\mathrm{e})$ is shown in Fig.~\ref{fig:magnetotransport}b. Below a threshold of $v_\mathrm{e}\sim 1.5$~m/s the DWs do not move. A linear behaviour is observed above this threshold, for which a straight line fit returns a slope of $11.6 \pm 0.3$. This linear behaviour, along with the order of magnitude of the velocities,  excludes the creep regime\cite{duttagupta2016} and confirms that all our data are measured in the viscous flow regime for domain wall motion\cite{metaxas}. The spin drift velocity $v_\mathrm{e}$ is the fastest possible DW velocity in the adiabatic limit, but $v > v_\mathrm{e}$ has been previously observed in single layer wires by Hayashi et al. \cite{Hayashi2007a}, for instance. This can be explained by the action of the nonadiabatic contribution to the spin transfer torque\cite{thiavilleepl2005}, leading to the expression $v = (\beta / \alpha) v_\mathrm{e}$ in the regime below Walker breakdown. This relation was also verified for the SAF and SyF structures using micromagnetic simulations and found to hold.

There is extensive data on both $\beta$ and $\alpha$ for Permalloy thin films, but much less is known about the CoFe alloy that we have used here. Whilst these quantities are likely to have similar orders of magnitude in these two transition metal magnets, it is unlikely that the values will be exactly the same. Vector-network-analyser ferromagnetic resonance measurements, described in more detail in the Supplementary Information, determined $\alpha = 0.007 \pm 0.001$ in our material, slightly less than that usually measured for Permalloy thin films\cite{Walowski2008,Lepadatu2010,Weindler2014}, but consistent with other measurements of CoFe alloys \cite{Rantschler2003,schoen2016}.

Combining this with the slope of the $v(v_\mathrm{e})$ line implies that $\beta = 0.08 \pm 0.01$ in our CoFe layers. Again, this is very similar to\cite{Eltschka2010} or just higher than\cite{Lepadatu2009a,Lepadatu2009b} typical values for Permalloy, justifying our assumption that nonadiabatic effects are important in this case. Values of $\beta$ in excess of 0.1 have been measured for vortex cores in Permalloy \cite{Heyne2010,Pollard2012,Roessler2014}, where the very high magnetisation gradients around the core can lead to enhanced local values for $\beta$\cite{ClaudioGonzalez2012}. The fact that our simple N\'{e}el DWs are narrower than typical transverse DWs in soft magnetic nanowires can therefore be expected to lead to a mild enhancement of $\beta$. Combined with the reduced value for $\alpha$, these two effects lead to a raised value for the $\beta/\alpha$ ratio.

It is noteworthy that the zero-field critical current density for the onset of motion of this 180\degrees~coupled DW pair is just smaller than $1.0 \times 10^{11}$~Am$^{-2}$, a remarkably low value compared to the typical values observed in single-layer Permalloy magnetic nanowires, which are in the $10^{12}$~Am$^{-2}$ range\cite{yamaguchiprl2004,Meier2007,Lepadatu2009a,Lepadatu2009a,Heyne2009}. It is even lower then the $2.5 \times 10^{11}$~A/m$^2$ threshold current density achieved in an optimised Co/Ni perpendicularly magnetised stack. It represents a roughly fivefold reduction over the critical current density in perpendicularly magnetised SAF nanowires\cite{Yang2015}, in which the DWs are driven by highly efficient spin Hall torques and giant exchange torques. In these in-plane SyFs, only the comparatively inefficient volume spin transfer torques, with an unremarkable value of the nonadiabaticity constant $\beta$ are able to depin the coupled DW pairs at this low value of current density, which has to be compared with a depinning field $H_\mathrm{depin} \sim 150$~Oe, taken from the magnetoresistance curves. At these low current densities around critical value, temperature rises due to Joule heating are limited to a few kelvins, meaning that thermal effects such as the nucleation of additional domain walls, can be safely ruled out.

Control measurements have been carried out on ferromagnetically coupled wires identical in every respect but for slightly thicker (1~nm) Ru spacer layers, changing the sign of the coupling. These wires have a similar anisotropic magnetoresistance response to their AF-coupled counterparts, allowing them to be tested for current-driven domain wall motion by the same method. As shown in Fig.~\ref{fig:magnetotransport}b, no domain wall motion was observed at zero field, in contrast to the antiferromagnetically coupled wires, before irreversible changes to the resistance occur for current densities exceeding about $8 \times 10^{11}$~A/m$^2$, directly showing the effect of AF coupling in a wall pair in reducing the critical current.

The threshold current density has been shown to vary with pulse duration\cite{fukami2013,parkin2014domain}, rising when pulses are very short. Our experiments use long pulses of a few hundred or thousand ns, but the pulse length dependence experiments in these previous reports show that our results are comparable with any others that use pulses that are longer than a few ns.

The DWs display a 2-D structure in microscope imaging. Notches and impurities that pin the walls are also 2-D objects. Therefore the dynamics of the walls are quantitatively affected by 2-D effects which are not captured by the 1-D model. However, in the 1-D simulations the calculated in-plane coupling at the point where the threshold current is removed completely is $\Delta_{\parallel}=-1.9$~mJ/m$^2$. This is of the same order as the estimated total interlayer coupling in experiments (see Methods) and gives evidence that the large reduction in the DW threshold current is due to nonadiabatic driving in the presence of highly anisotropic interlayer coupling. The fact that the experimental depinning current density is slightly below that predicted by the model is due to the fact that the wall is not in reality a completely rigid object, as can be seen from Fig.~\ref{fig:SAF_LTEM}c for instance, which allows it to be slightly more easily depinned. Simulations using the 1D model (described in the Supplementary information) show that the reduction in critical current density is expected to be even larger in a properly balanced SAF.

\section*{Discussion}

Here we have shown that the magnetostatics in in-plane magnetised SyF nanowires lead to coupled pairs of simple N\'{e}el walls that are narrow, rigid and unaffected by the wire width in which they are present. The mixture of exchange and magnetostatic coupling between the walls in the pair leads to an anisotropic coupling overall, which even has different signs in the in-plane and out-of-plane directions. This anisotropic coupling, when combined with non-adiabatic driving, leads to very effective depinning from an extrinsic pinning potential by exploiting the internal degrees of freedom within the coupled wall pair that do not exist in a single wall. We realised this experimentally at the very low current density of $10^{11}$~Am$^{-2}$, roughly one order of magnitude lower with respect to typical single layer ferromagnetic wires, and at least an eightfold reduction when compared to a control sample that was identical in every way but for being ferromagenetically coupled through a slightly thicker Ru spacer. Since power dissipation goes quadratically with current, significant reductions in power consumption can be achieved by properly exploiting this phenomenon.

Although we studied the nature of these internal degrees of freedom and how they may be exploited in the setting of in-plane magnetised materials with the domain walls driven by volume spin-transfer torques, these ideas are not limited to that context. The important feature of the non-adiabatic component of the volume spin-transfer torque is that it has a field-like symmetry. Such field-like components are also present in interfacial spin-orbit torques \cite{garello2013}, which tend to be very efficient. We can therefore expect that our results can be used to design appropriately configured coupled DW pairs in spin-orbit torque-driven systems in order to realise a similar reduction in the current density needed for depinning. Thus, these systems show great promise for future devices based on current-driven DW motion that are very energy efficient.

\section*{METHODS}

\footnotesize{

\noindent \textbf{Sample Fabrication and Measurement.} Thin films of substrate / Ru~(2) / Co$_{90}$Fe$_{10}$~(10) / Ru~(0.7) / Co$_{90}$Fe$_{10}$~(10) / Ru~(2) -- SAF, thickness values in nm -- and susbstrate / Ru~(2) / Co$_{90}$Fe$_{10}$~(6.6) / Ru~(0.7) / Co$_{90}$Fe$_{10}$~(13.3) / Ru~(2) –- SyF –- were sputtered using ultrapure Ar gas on thermally oxidized Si substrates for XMCD-PEEM imaging, thin-film characterisation, and electrical measurements, and on Si$_3$N$_4$ membranes for LTEM imaging. The layer closest to the sample surface is the top layer with thickness $t_1$, whilst the bottom layer closest to the susbtrate has thickness $t_2$. Samples were patterned by electron-beam lithography and an Al/Ti hard-mask was sputtered and obtained by lift-off. The wire width was 400~nm for PEEM and magnetotransport measurements, and 800~nm for LTEM. Ar-ion milling was used to remove the thin film not covered by the Al/Ti hard-mask and the final patterns were obtained by etching the Al layer using MF319, removing the Al/Ti hard-mask. Electrical contacts were obtained by optical lithography and sputtering of Ru~(5~nm) / Au~(80~nm). Magnetoresistance measurements were carried out at room temperature using a lock-in amplifier method at 10~kHz, with 1~$\upmu$A amplitude source current. The rise/fall time of the current pulses used to measure the wall velocity was $\sim 5$~ns.

The magnetic and structural properties of the unpatterned samples were investigated using VSM, x-ray diffraction and TEM imaging using 5~mm~$\times$~5~mm thin-films sputtered on thermally oxidized Si substrates. The saturation magnetisation $M_\mathrm{s}$ of the Co$_{90}$Fe$_{10}$ layers was measured to be 1.40~MA/m. The measured saturation field implies a bilinear interlayer indirect exchange coupling constant \[J_1 = -\frac{\mu_0 M_\mathrm{S}}{2} \left( \frac{2 t_1 t_2}{t_1 + t_2} \right)  H_\mathrm{S} = -1.0~\mathrm{mJ/m}^2, \] where $H_\mathrm{s}$ is the saturation field of the SyF structure. The spin-polarisation, $P$, of the Co$_{90}$Fe$_{10}$ layers, was $0.34 \pm 0.02$ as determined from spin-wave Doppler measurements on a Ru/Co$_{90}$Fe$_{10}$/Ru trilayer in the magnetostatic surface wave geometry\cite{zhu2010,sugimoto2016}. The current density, $j$, used in the experiments was converted to spin drift velocity, $v_\mathrm{e}$, as $v_\mathrm{e} = Pg\mu_\mathrm{B}J/2eM_\mathrm{S}$.

\noindent \textbf{Magnetic Imaging.} XMCD-PEEM imaging was carried out at the I06 beamline of the Diamond Light Source synchrotron, using circularly-polarised photons tuned the Co L$_3$ edge at 779~eV. The samples were mounted on a PEEM cartidge containing a small electromagnet with maximum field at the sample of $\pm 250$~Oe. All images were obtained at remanence. The x-ray beam is directed along the track and the magnetisation projection along the beam direction is shown as a gray-scale. The escape depth of the photelectrons is only a few nm and thus only the top Co$_{90}$Fe$_{10}$ layer contributes any significant magnetic contrast.

The Lorentz imaging experiments described here were performed on a Tecnai~T20 TEM. In this mode the  objective lens of the microscope is weakly excited to provide a field perpendicular to the thin film sample and a magnetising sequence is then carried out by tilting the sample in this field\cite{mcvitie2015}. Magnetic contrast is generated by defocussing the image forming lens which reveals DWs as bright and dark lines on a uniform background. Furthermore in the case of magnetic nanowires similar contrast can be observed along the wire edges where the magnetisation has a component parallel to the edge. The magnetisation reversal of the structure was carried out in Lorentz TEM mode with a weak excitation of the objective lens set to a  value of 1100~Oe and the sample then tilted so that an in-plane field along the wire axis of 650~Oe was applied, which is enough the saturate both layers in the field direction, as is shown in the starting state in Fig.~\ref{fig:SAF_LTEM}a.

\noindent \textbf{1-D Model of Domain Wall Dynamics.} The 1-D model of SAFs has been developed in Ref.~\onlinecite{Saarikoski2014} and involves a system of N{\'e}el-type DWs in a synthetic antiferromagnet where magnetisation is in the plane of the wire. The walls are described using two parameters for each wall; the positions of the walls along the wire $Z_i$, $i=1,2$ and the angles of the spins with respect to the wire plane at the center of the wall $\phi_i$, $i=1,2$ (see Fig. \ref{fig:NONADIABATIC_DRIVING}a). The model incorporates interlayer coupling terms due to magnetostatic effects and the indirect exchange interaction between the layers. The total coupling is ferromagnetic for the out-of-plane component $\Delta_{\perp}$ and antiferromagnetic for the in-plane coupling $\Delta_{||}$. In view of the coupling parameters obtained from micromagnetic simulations we use a ratio $\Delta_{\perp}/\Delta_{||}=-4$ in the 1-D simulations, \textit{i.e.} a significantly stronger out-of-plane coupling. For fuller details of the 1-D model and the equations of motion, see the Supplementary Information.

We assumed a Co-like value for the exchange stiffness $A = 30$~pJ/m. Demagnetising factors for the layers were calculated using the formulae given by Aharoni \cite{Aharoni1997}, leading to the following shape anisotropy parameters: $K_{x,1} = 63.9$~kJ/m$^3$, $K_{y,1} = 1.17$~MJ/m$^3$, $K_{x,2} = 36.2$~kJ/m$^3$,$K_{y,2} = 1.20$~MJ/m$^3$. This leads to wall width parameters $\Lambda_i = \sqrt{A/K_{x,i}}$, \text{i.e.} neglecting any effects of coupling, of $\Lambda_1 = 22$~nm and $\Lambda_2 = 29$~nm. Note that this wall width parameter is not necessarily the same as the apparent DW width in \textit{e.g.} an LTEM image\cite{Beach2008}. We used a Gilbert damping parameter $\alpha=0.007$ (measured by ferromagnetic resonance as described in the Supplementary Information) and nonadiabatic torque constant $\beta=0.08$ (measured from the slope of the $v(v_\mathrm{e})$ curve in Fig.~\ref{fig:magnetotransport}b). The pinning potential is of the extrinsic type and is modelled with a parabolic potential well in one of the layers with a strength equivalent to a 150 Oe field, equivalent to the depinning field observed in experiments. The initial conditions for the walls correspond to the ground state with both walls at a value of $Z$ equal to that at the center of the pinning potential.

\noindent \textbf{Micromagnetic Modelling.} Micromagnetic simulations were performed using the \textsc{boris} micromagnetics software developed by us. The simulations were ran using the finite difference method and the LLG and LLG-STT equations were solved explicitly using the 2nd order Adams-Bashforth-Moulton method with adaptive time-step. Exchange fields were calculated using the 6-neighbour stencil with Neumann boundary conditions and magnetostatic fields were calculated using FFT-based convolution with zero padding. The mesh cell size was 5~nm. The in-plane and out-of-plane coupling parameters were calculated respectively by obtaining the difference between the total energy density values, including magnetostatic, direct exchange and bilinear surface exchange interactions, for anti-parallel and parallel N\'{e}el DW rotations in the two layers respectively. For the in-plane coupling $\Delta_{||}$ the N\'{e}el wall rotation occurs in the plane, whilst for the out-of-plane coupling $\Delta_{\perp}$ the N\'{e}el wall rotation occurs perpendicular to the plane. When combined with the isotropic bilinear indirect exchange coupling $J_1 = -1.0$~mJ/m$^2$, we obtained overall couplings of $\Delta_{||} = -0.75$~mJ/m$^2$ and $\Delta_{\perp} = +3.02$~mJ/m$^2$.}

\bibliographystyle{naturemag}

\bibliography{SAF}

\vskip 2cm

\section*{Acknowledgments}
This work was partly supported by the EPSRC through grants EP/I011668/1 and EP/I011668/1, the Marie Curie International Incoming Fellowship SKYHIGH, and by a Grant-in-Aid for Scientific Research (C) No. 26390014 from Japan Society for the Promotion of Science. We thank Diamond Light Source for the provision of beamtime under proposal number SI-6709.

\section*{Author contributions}
S.L., T.A.M., G.B., D.M., S.M, and C.H.M. planned the experiments. H.S., G.T., and C.H.M. planned the theoretical calculations, which were undertaken by H.S. S.L. fabricated the samples and performed magnetotransport measurements. D.Y. grew the samples for and performed the FMR measurements with the assistance of M.C.W. S.S. prepared the samples for and performed the spin-wave Doppler measurements. R.B. fabricated the Lorentz TEM specimens and performed the imaging and image analysis. M.J.B. fabricated the TEM cross-section specimens and performed the imaging and image analysis. S.L., R.B, M.J. B., T.A.M., G.B, D.M., S.M., J.M., and S.S.D. performed the PEEM imaging, S.L. performed the image analysis. S.L. developed the bespoke micromagnetics code \textsc{boris} and performed the micromagnetic modelling. S.L., H.S., and C.H.M. wrote the paper. All authors discussed the data and reviewed the manuscript.

\section*{Additional information}
\textbf{Supplementary Information} accompanies this paper at http://www.nature.com/srep

\textbf{Competing financial interests:} The authors declare no competing financial interests.

\textbf{Reprints and permission} information is available online at http://npg.nature.com/reprintsandpermissions/

\newpage

\end{document}